\documentclass{jfm}
\usepackage{graphicx}
\usepackage{epstopdf, epsfig}
\usepackage{natbib}
\usepackage{graphics}
\usepackage{amsmath}
\usepackage{amssymb}
\usepackage{mathrsfs}
\usepackage{eucal}		
\usepackage{wasysym}
\usepackage{caption}
\usepackage{subcaption}
\usepackage{multirow}
\usepackage[hidelinks]{hyperref}

\shorttitle{Deep learning of mixing by two `atoms' of stratified turbulence}
\shortauthor{H. Salehipour, W. R. Peltier}

\title{Deep learning of mixing by two `atoms' of stratified turbulence}

\author{Hesam Salehipour\aff{1,2}
  \corresp{\email{h.salehipour@utoronto.ca}},
  W. R. Peltier\aff{1}}

\affiliation{\aff{1}Dept. of Physics, University of Toronto, Toronto, ON M5S 1A7, Canada
\aff{2}Autodesk Research, MaRS Discovery District, 661 University Ave, Toronto, ON M5G 1M1, Canada}

\begin{document}

\maketitle

\begin{abstract}
Current global ocean models rely on ad-hoc parameterizations of diapycnal mixing, in which the efficiency of mixing is globally assumed to be fixed at 20\%, despite increasing evidence that this assumption is questionable. As an ansatz for small-scale ocean turbulence, we may focus on stratified shear flows susceptible to either Kelvin-Helmholtz (KHI) or Holmboe wave (HWI) instability. Recently, an unprecedented volume of data has been generated through direct numerical simulation (DNS) of these flows. In this paper, we describe the application of deep learning methods to the discovery of a generic parameterization of diapycnal mixing using the available DNS dataset. We furthermore demonstrate that the proposed model is far more universal compared to recently published parameterizations. We show that a neural network appropriately trained on KHI- and HWI-induced turbulence is capable of predicting mixing efficiency associated with unseen regions of the parameter space well beyond the range of the training data. Strikingly, the high-level patterns learned based on the KHI and weakly stratified HWI are `transferable' to predict HWI-induced mixing efficiency under much more strongly stratified conditions, suggesting that through the application of appropriate networks, significant universal abstractions of density stratified turbulent mixing have been recognized.
\end{abstract}
\vspace{-1.5cm}

\section{Introduction} 
\label{sec:intro} 
\begin{figure}
  \centerline{\includegraphics[width=.8\textwidth]{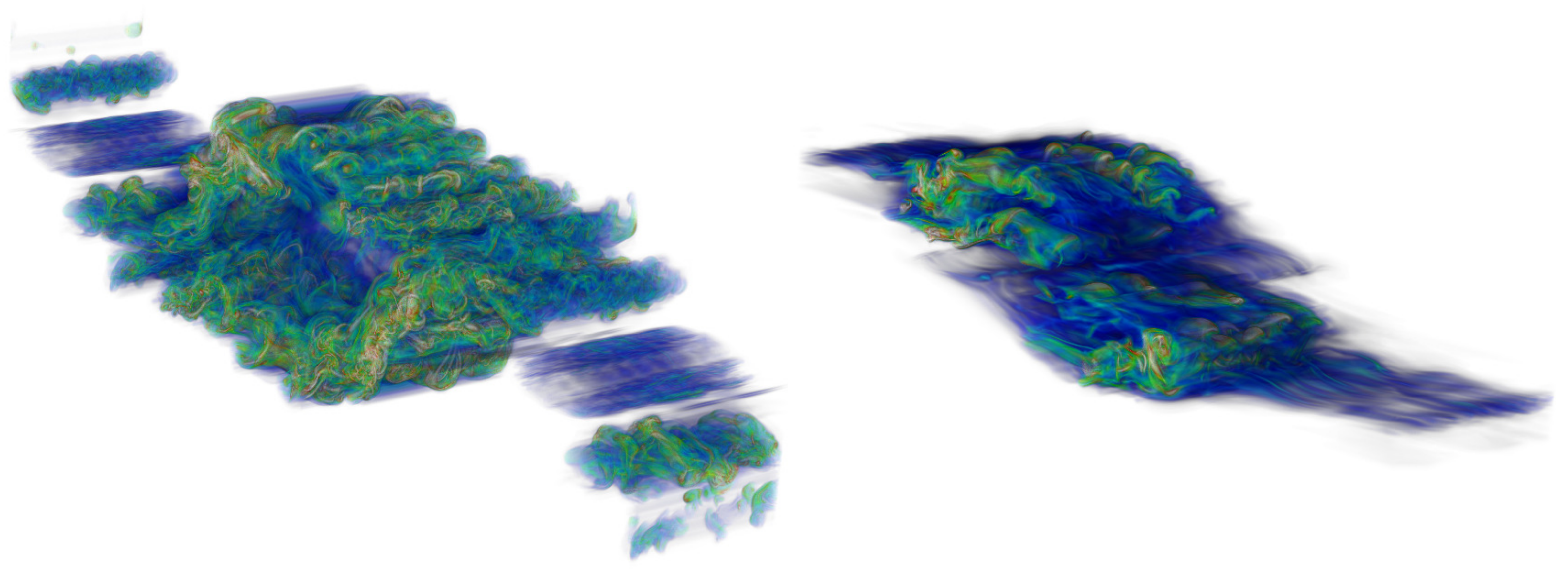}}
  \caption{Schematic of two `atoms' of turbulence in stratified shear flows associated with Kelvin-Helmholtz instability (left) and Holmboe instability (right).}
\label{fig:pretty}
\end{figure}

%
A vital mechanism for ventilating the abyssal ocean is that due to vertical mixing of deep, cold and nutrient-rich waters with shallower, warm and nutrient-scarce waters \citep{Wunsch_Ferrari_2004}. Mediated by the complex interactions of the internal wave field in the ocean interior, these mixing events emerge at the smallest scales and undergo transition to turbulence that leads to an irreversible conversion of kinetic energy to potential energy. Figure \ref{fig:pretty} demonstrates two flavors of these events that may develop in stratified shear flows, namely the Kelvin-Helmholtz instability (KHI) and the Holmboe wave instability (HWI) (refer to \cite{SCP16, SOC_Hesam} for an in-depth comparison of these instability mechanisms). 
Despite their critical role in modulating the large scale meridional overturning circulation of the ocean, the effect of these small scale `atoms' of ocean turbulence are often overly simplified by parameterizing them as involving a constant mixing rate that is always $20\%$ of the local dissipation rate of kinetic energy \citep{Gregg_etal_2018_review}. However, detailed numerical simulations and experimental measurements have collectively demonstrated significant departures from this fixed canonical value (see \eg \cite{Monismith_GRL_2018}). Recent advances have been made in proposing alternative parameterizations of mixing \emph{efficiency} based on forced and homogeneously stratified flows (see \eg \cite{mater2014quest, Maffioli2016mixing}) or freely-evolving and inhomogeneously stratified flows (see \eg \cite{Hesam_GRL, Mashayek_etal_2017_GRL}). Even in the latter conditions that are more realistic, the focus has been mainly on the fully turbulent flows for which the imprint of the initial `atom' involved is minimal. For instance the effect of ubiquitous large overturns (see \eg figure \ref{fig:pretty}) that are convectively unstable leading to highly efficient mixing (with efficiency as high as 0.8-0.9) have been ignored in these earlier investigations.

Our main goal in this paper is to propose a data-driven approach that substantially improves previous parameterizations by encompassing \emph{all} the data that is available based on direct numerical simulation of these `atoms'. To introduce this approach in the current study we focus on two of the distinct archtypical flavors of stratified turbulence. Section \S\ref{sec:dataset} presents the cornerstone of this paper that involves a large compilation of data associated with KHI and HWI. These data are prepared in the manner described in \S\ref{sec:pre_process} to be further analyzed in \S\ref{sec:cnn} based on the application of `deep learning' methods. We evaluate the predictions of this data-driven approach and compare them with previous methods in \S\ref{sec:results}. Our findings and discussion of future research directions are summarized in \S\ref{sec:summary}.
%


\section{The Parent DNS dataset} 
\label{sec:dataset} 

\begin{table}
  \begin{center}
\def~{\hphantom{0}}
  \begin{tabular}{lccccc|ccccc}
  \multicolumn{6}{c}{Training dataset} & \multicolumn{5}{|c}{Validation dataset}\\
  \hline
      \multirow{24}{*}{KHI} &
      $Re$   & $Ri_b$	&   $Pr$ & $R$ 	& $n_s$ & $Re$   & $Ri_b$ &   $Pr$ & $R$ & $n_s$ \\[3pt]
      & ~6000  & 0.12~	&   1	 & 1   & 201 & ~~6000   & 0.12~ &   16	 & 1 &  298 \\
      & ~6000  & 0.12~	&   2	 & 1   & 250 & ~~6000   & 0.001 &   1	 & 1 &	172 \\
      & ~6000  & 0.12~	&   4	 & 1   & 200 & ~~6000   & 0.22~ &   1	 & 1 & 	185 \\
      & ~6000  & 0.12~	&   8	 & 1   & 188 & ~20000   & 0.16~ &   1	 & 1 &	150 \\
      & ~6000  & 0.005	&   1	 & 1   & 409 & ~30000   & 0.12~ &   1	 & 1 &	169 \\ 
      & ~6000  & 0.01~	&   1	 & 1   & 541 \\       
      & ~6000  & 0.02~	&   1	 & 1   & 293 \\       
      & ~6000  & 0.04~	&   1	 & 1   & 208 \\       
      & ~6000  & 0.08~	&   1	 & 1   & 100 \\       
      & ~6000  & 0.16~	&   1	 & 1   & 150 \\       
      & ~6000  & 0.20~	&   1	 & 1   & 150 \\
      & ~6000  & 0.02~	&   8	 & 1   & 126 \\       
      & ~6000  & 0.04~	&   8	 & 1   & 133 \\       
      & ~6000  & 0.10~	&   8	 & 1   & 150 \\       
      & ~6000  & 0.14~	&   8	 & 1   & 125 \\       
      & ~6000  & 0.16~	&   8	 & 1   & 150 \\       
      & ~6000  & 0.18~	&   8	 & 1   & 154 \\       
      & ~6000  & 0.20~	&   8	 & 1   & 146 \\ 
      & ~4000  & 0.16~	&   1	 & 1   & 150 \\
      & ~4000  & 0.16~	&   8	 & 1   & 150 \\   
      & ~8000  & 0.16~	&   1	 & 1   & 150 \\
      & ~8000  & 0.16~	&   8	 & 1   & 106 \\
      & 12000  & 0.16~	&   1	 & 1   & 150 \\      
      \hline\multirow{8}{*}{HWI}&
      $Re$   & $Ri_0$	&   $Pr$ & $R$ 	& $n_s$&  $Re$   & $Ri_0$ &   $Pr$ & $R$ & $n_s$ \\[3pt]
      & ~4000  & 0.16  & 8 & 2.83 & 250 & ~~6000  & 0.32  & 8 & 10   & 183 \\
      & ~6000  & 0.16  & 8 & 2.83 & 201 & ~~6000  & 0.32  & 8 & 5    & 267 \\ 
      & ~6000  & 0.16  & 8 & 5    & 195 & ~~6000  & 0.16  & 8 & 25   & 187 \\     
      & ~6000  & 0.16  & 8 & 10   & 163 \\
      & ~6000  & 0.08  & 8 & 5    & 214 \\     
      & ~6000  & 0.08  & 8 & 10   & 182

%
  \end{tabular}
  \caption{The collection of initial parameters (as defined in equation \eqref{eq:params}) employed for conducting DNS experiments associated with either KHI or HWI. $n_s$ indicates the number of saved snapshots for each individual simulation. The split between training and validation sets are also highlighted.}
  \label{tab:siminfo}
  \end{center}
\end{table}

We model a stratified mixing layer by assuming initial velocity and density distributions that have hyperbolic tangent form, as:
\begin{equation}
  \overline{u}(z,0)= U_0 \tanh \left (\frac{z}{d} \right ), 
  \quad
  \overline{\rho} (z,0) = \rho_0 \left[ 1 -  \tanh   \left (\frac{z}{\delta} \right ) \right], 
  \label{eq:tanhprof}
\end{equation}
in the Boussinesq approximation such that $\rho_0 \ll \rho_r$ (note that here density represents departures from a hydrostatic state associated with $\rho_r$). Also, $U_0$ and $\rho_0$ denote respectively half the total velocity and density jumps across the shear layer (with a total depth $2d$) and the density layer (with a total depth of $2\delta$). As a result of this canonical setting, the dimensionless Boussinesq equations are governed by four important non-dimensional parameters, namely the (initial) Reynolds number $Re$; the bulk Richardson number $Ri_b$; the Prandtl number $Pr$; and the initial scale ratio $R$, defined altogether as:
\begin{equation}
Re = \frac{U_0 d}{\nu}, 
\quad
Ri_b=\frac{g \rho_0 d}{\rho_r U_0^2},
\quad 
Pr = \frac{\nu}{\kappa},
\quad
R=\frac{d}{\delta}, 
\label{eq:params}
\end{equation}
in which $\nu$ is the kinematic viscosity, $\kappa$ is the molecular diffusivity and $g$ is the gravitational acceleration. 
Table \ref{tab:siminfo} lists all the DNS analyses from which data will be employed for training and validation of the proposed artificial neural networks. These simulations have been thoroughly analyzed and discussed previously in a number of recent publications on KHI \citep{SPM15, SP15, Hesam_GRL} and HWI \citep{SCP16, SOC_Hesam}. For details of each simulation, interested readers are referred to the relevant paper. 

For the supervised machine learning application to be discussed herein, we have further subdivided these datasets into \emph{training} and \emph{validation} sets with an approximate 80\%-20\% ratio, as indicated in table \ref{tab:siminfo}. Both these subsets include examples of flow evolution due to KHI and HWI. We have intentionally chosen the validation dataset to include all DNS cases with extreme values for their initial parameters, that are well beyond the range of similar parameters employed for training purposes. This enables us to investigate the extent to which our trained model is generalizable and thus robust. Note that our training dataset had a very limited number of HWI examples (compared to KHI) that are also at much smaller values of $Ri_b$ and $R$.

\section{Pre-processing of DNS data} 
\label{sec:pre_process} 
The results of each three-dimensional DNS experiment associated with the evolution of either KHI or HWI, is comprised of $n_s$ snapshots in time where each saved snapshot represents three-dimensional fields of flow quantities, namely the density $\rho$ and velocity fields $\boldsymbol{u} = (u,v,w)$ (Table \ref{tab:siminfo} lists $n_s$ for each simulation). The intensity of turbulent activity may be represented by the pointwise dissipation rate of total kinematic energy, $\epsilon(\boldsymbol{x},t)$  defined as,
\begin{equation}
  \epsilon (\boldsymbol{x},t) = 2\nu s_{ij}s_{ij},
  \label{eq:epsilon}
\end{equation}
in which $s_{ij} = \left( \partial u_i / \partial x_j + \partial u_j / \partial x_i \right)/2 $ is the total strain rate tensor. We may also reduce the above three-dimensional fields into a one-dimensional profile by performing horizontal averaging (to be denoted here by an overbar). Thus the horizontally-averaged dissipation rate of total kinematic energy, $\overline{\epsilon} (z,t)$, and the mean flow density, $\overline{\rho}(z,t)$, are defined as, 
\begin{equation}
  \overline{\epsilon} (z,t) = \frac{1}{L_xL_y}\displaystyle\int \epsilon(\boldsymbol{x},t) \ dxdy,
  \qquad 
  \overline{\rho}(z,t) = \frac{1}{L_xL_y}\displaystyle\int \rho(\boldsymbol{x},t) \ dxdy,
  \label{eq:epsilon_z}
\end{equation}
where $L_x$ and $L_y$ denote the size of the computational domain in the streamwise and spanwise directions.

The (generally) time dependent \emph{mixing efficiency}, $\mathscr{E}$, may be computed precisely by invoking the concept of irreversible diapycnal mixing (originally introduced by \citet{Winters_1995}) which relies on a special kind of reduction operator, namely a three-dimensional sorting of the density field into a notional state that is strictly stably stratified \citep{PC03}, and is defined as
\begin{equation}
 \mathscr{E}(t) = \frac{\mathscr{M}(t)}{\mathscr{M}(t) + \langle \overline{\epsilon}(z,t) \rangle},
 \label{eq:eff_dns}
\end{equation}
where $\langle \rangle$ denotes vertical averaging. For a precise definition of $\mathscr{M}(t)$ refer to equation (2.18) of \cite{SCP16} and the cited discussions therein. The required parallel implementation of the sorting procedure is described in \cite{SPM15} (see \eg their figure 1). Such an elaborate technique for calculating $\mathscr{E}$ is only viable in numerical simulations such as those employed in this work because in practice oceanographers only measure one-dimensional profiles in depth and are therefore unable to perform the same analysis. Indeed, there is a similar subtlety in defining the `background' buoyancy frequency, $N^2(z,t)$ as described in \cite{SP15} (see their discussion leading to equation (2.23)) and more recently in \cite{Arthur_etal_2017}. In order to distinguish between irreversible mixing and reversible stirring, $N^2(z,t)$ must be defined based on the same notional state obtained by the three-dimensional sorting procedure. For consistency with common practice in oceanography, in this paper we may define $N^2$ using the mean flow density introduced in \eqref{eq:epsilon_z} such that $N^2(z,t) = -(g/\rho_r){d\overline{\rho}/dz}$.

We seek a mapping between the instantaneous vertical profiles of $\overline{\epsilon}(z,t_0)$ and $N^2(z,t_0)$ (i.e. at a given time $t_0$) and the precisely computed values of mixing efficiency, $\mathscr{E}(t_0)$. Once the network is trained, this mapping would essentially reveal a reduction operator that is conceivably very different from a straightforward vertical averaging, one which also incorporates the structural pattern and length scales that implicitly exist and are thus `hidden' in these profiles. Thus the inputs to our artificial neural network are tuples of $(\mathcal{X}_1, \mathcal{X}_2)$ defined respectively as,
\begin{equation}
  \mathcal{X}_1(z,t_0) \equiv \frac{\overline{\epsilon} (z,t_0)}{\kappa \displaystyle\int N^2(z,t_0) \ dz}, 
  \qquad 
  \mathcal{X}_2(z,t_0) \equiv \frac{N^2(z,t_0) }{\displaystyle\int N^2(z,t_0) \ dz}.
  \label{eq:inputs}
\end{equation}
Furthermore, the true `labels' in our supervised learning setting are the instantaneous values of mixing efficiency, namely
\begin{equation}
  \mathcal{Y}(t_0) \equiv \mathscr{E}(t_0).
  \label{eq:labels}
\end{equation}
It is important to highlight that $(\mathcal{X}_1, \mathcal{X}_2)$ appear in a normalized form to render $\overline{\epsilon}(z,t_0)$ and $N^2(z,t_0)$ comparable in terms of their dimensionality and physical relevance and furthermore to extend the applicability of the trained network to oceanographic profiles. In equation \eqref{eq:inputs}, $\mathcal{X}_1$ represents the vertical profile of kinetic energy dissipation rate relative to the molecular diffusion rate in the absence of mean flow shear. 
The vertical profiles are assumed to have a fixed length of $i=512$ points in which the `dead' regions of the simulation (near top and bottom boundaries) have been excluded by focusing on the largest segments of the profiles where $|d\overline{\rho}(z)/ dz| \geq 10^{-3}$. This approach is analogous to identifying `patches' of turbulence from DNS calculations \citep{Smyth_etal_2001}.


\section{Deep Convolutional Neural networks} 
\label{sec:cnn} 

Deep learning methods involve a multilayer stacking of simple modules that perform linear or nonlinear input-output mappings whose weights and biases are subject to `training' through an optimization procedure \citep{lecun2015_nature}. These techniques became widely popularized after \cite{Krizhevsky_etal_2012}, from University of Toronto, employed a `deep convolutional neural network' to classify a dataset of 1.2 million images and won the first place in the 2012 ImageNet competition. A convolutional neural network (CNN) is a special type of neural network architecture that relies on the convolution operator in lieu of general matrix multiplication in at least one layer of its configuration \citep{DeepLearning_book}. In two dimensions, this operator may be defined as,
\begin{equation}
 \widetilde{\boldsymbol{\mathcal{X}}}(i,j) = (\boldsymbol{\mathcal{X}} \ast \mathscr{K} )(i,j) = \sum_m\sum_n \boldsymbol{\mathcal{X}}(m,n) \mathscr{K}(i-m, j-n)
 \label{eq:convolution}
\end{equation}
where the input field $\boldsymbol{\mathcal{X}} \in \mathbb{R}^{i\times j}$ and the convolution kernel $\mathscr{K} \in \mathbb{R}^{m\times n}$ is represented by characteristic filter lengths of size $m \leq i$ and $n \leq j$. A convolved (i.e filtered) field is constructed by traversing the kernel $\mathscr{K}$ over the dimensions of $\boldsymbol{\mathcal{X}}$. To keep the filtered field, $\widetilde{\boldsymbol{\mathcal{X}}}$, the same size as $\boldsymbol{\mathcal{X}}$ often zero-padding is employed.

\begin{figure}
  \centerline{\includegraphics[width=0.8\textwidth]{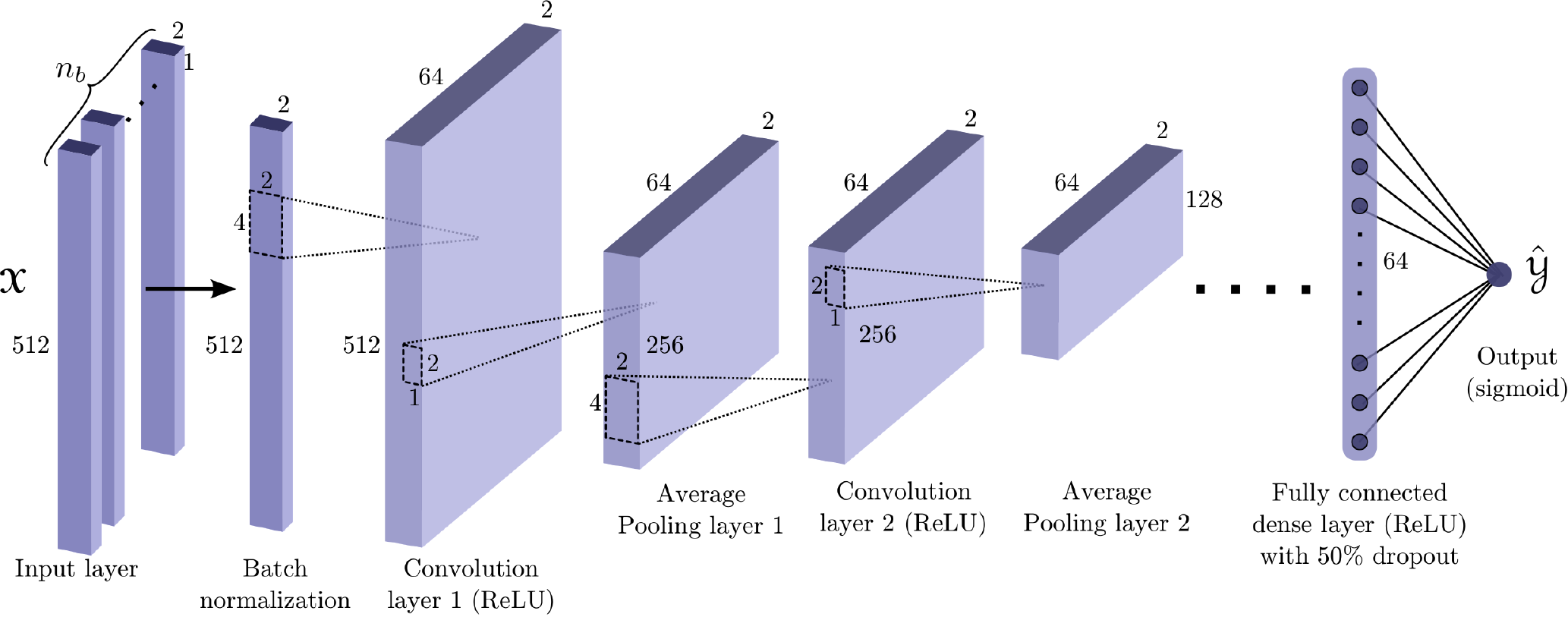}}
  \caption{Illustration of the type of convolutional neural network investigated in this paper with increasing number of stacked layers. Refer to the text for definition of various layers. For any given example data point, the tuples of $\boldsymbol{\mathcal{X}}= (\mathcal{X}_1, \mathcal{X}_2)$ are of size ($512\times2$) representing vertical profiles of normalized $\epsilon(z)$ and $N^2$ as defined in \eqref{eq:inputs}.}
\label{fig:cnn}
\end{figure}

Figure \ref{fig:cnn} illustrates the schematic configuration of the selected neural network architectures to be employed in this paper that may consist of one to seven convolution layers labeled as CNN1 to CNN7. Each configuration receives the input $\boldsymbol{\mathcal{X}} = (\mathcal{X}_1,\mathcal{X}_2) \in \mathbb{R}^{512\times2}$, as defined in \eqref{eq:inputs}, and passes it to a `batch normalization' layer \citep{batch_normalization} that, for any given `batch' of the training dataset (with size $n_b$), normalizes $\mathcal{X}_1$ and $\mathcal{X}_2$ individually by subtracting the batch mean and dividing by its variance. The batch-normalized data are then subsequently fed into a series of convolution layers each having 64 filters with a kernel $\mathscr{K}\in \mathbb{R}^{4\times2}$. Each convolution kernel undergoes a nonlinear activation function of type $f(x) = max(0,x)$, also known as a Rectified Linear Unit (ReLU). The output of each convolution layer is followed by an `average pooling' operator which effectively reduces the size of the profile by half through averaging any two adjacent data in the profile (i.e. averaging window of size $(2\times1)$).  The reduced outputs are then reshaped appropriately to be fed into a dense (or fully connected) layer with 64 neurons that also employs ReLU as its nonlinearity function.  To avoid overfitting and to improve the model predictions, we regularize the network by a method known as `dropout' \citep{Hinton2012_dropout} which randomly turns off 50\% of the neurons thereby eliminating their contribution in the `backpropagation' procedure (a method to apply the chain rule to derive gradients of the loss function with respect to trainable parameters in the network) of the optimization procedure. As a result the network is forced to learn robust features that emerge more frequently in the random subsets during training. Finally we use a single neuron to represent the network output, $\hat{\mathcal{Y}}$. We have chosen a sigmoid activation function for the output layer because efficiency values must be within 0 and 1. We have used the Adam optimizer \citep{AdamOptimizer} for performing stochastic gradient descent to minimize the loss function defined as the mean-squared-error $\sum_{n_b} (\mathcal{Y}-\hat{\mathcal{Y}})^2/n_b$ where the batch size is set as $n_b=100$.

\begin{figure}
  \centerline{\includegraphics[width=0.95\textwidth]{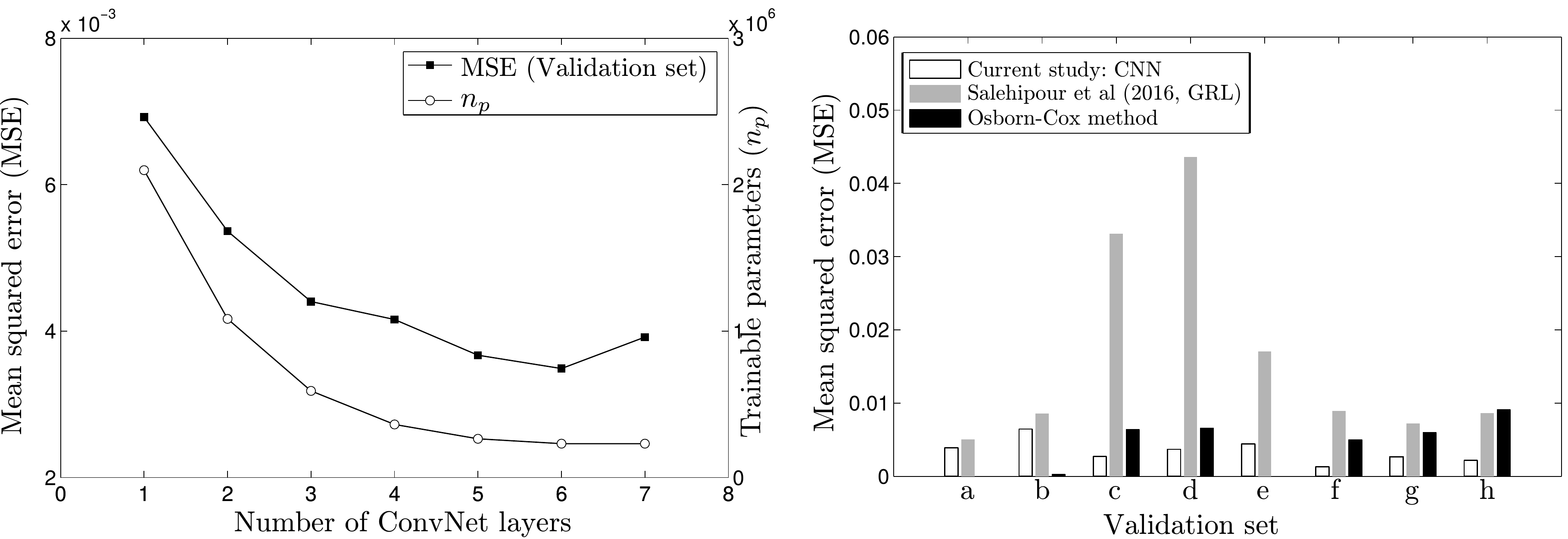}}
  \caption{Left: Comparing the predictions of CNN1-CNN7 on validation set. Right: the results of CNN6 (this study) are compared quantitatively with other methods in terms of mean squared error for various sets of validation data, labeled (a-h) as per figure \ref{fig:dl_eff}.}
\label{fig:dl_validationError}
\end{figure}

Notice that the number of trainable parameters, $n_p$, in the network decreases from CNN1 to CNN6 and increases slightly from CNN6 to CNN7 (see figure \ref{fig:dl_validationError}a). Taking CNN1 for instance, the network has copious neurons that link the first convolution layer (after pooling) to a fully connected dense layer leading to $n_p> 2 \times 10^6$. As the network becomes deeper, increasingly more structure is built into the network due to the locality of the convolution operators that may be contrasted to the global connectivity of the dense layers. By definition \eqref{eq:convolution}, the convolution layer shares its kernel parameters ($m\times n$ weights for a 2D kernel) across a given `tiling' of its input field. Notice that for deeper convolution layers whose inputs are derivative of the previous pooling operator with $q$ filters, there are $m\times n\times q$ trainable weights and one trainable bias that are shared by each tiling of the convolution layer. For instance in CNN2, $n_p \sim 64\times(4 \times 2+1)_{\small\mbox{conv1}} + 64\times (4\times 2\times 64+1)_{\small\mbox{conv2}} + (128 \times 2 \times 64 \times 64)_{\small\mbox{dense}}$. Refer to \citet[Chapter~9]{DeepLearning_book} for further details on \emph{parameter sharing} in convolutional networks. 
Figure \ref{fig:dl_validationError}a also evaluates the effect of increasing the depth of the network in so far as the validation data is concerned. Clearly CNN6 outperforms others which might be explained by the observed saturation of network training capacity also shown in this figure.

Obviously there are many parameters (or hyper-parameters) that we have assumed to be fixed within the above networks. Moreover, there are many other types of deep neural networks (DNN) \citep{DeepLearning_book} that could be exploited, an alternatively good candidate being the recurrent neural network which enables handling input sequences of vertical profiles with varying dimensions. We only note in passing that we also investigated a deep feed-forward neural network (that consists of deep stacking of dense layers only) in this work but the CNN results were substantially more accurate. We wish to emphasize that our focus in this paper has \emph{not} been to find the optimal configuration (or architecture) for producing the least possible error on the validation set.  This paper rather intends to make the first step in introducing the idea of employing deep learning methods for the purpose of parameterizing sub-grid scale processes using DNS datasets.
We have open-sourced our code and post-processed dataset in the hope of encouraging the community to further enhance such a data-driven approach to parameterization. 
Indeed, we believe the ultimate success of these efforts will rely upon a cohesive community-driven collaboration. Fortunately, these data-driven ideas are being embraced most recently in the boarder context of earth system modeling \citep{schneider2017}.

\section{Results} 
\label{sec:results} 

\begin{figure}
  \centerline{\includegraphics[width=.9\textwidth]{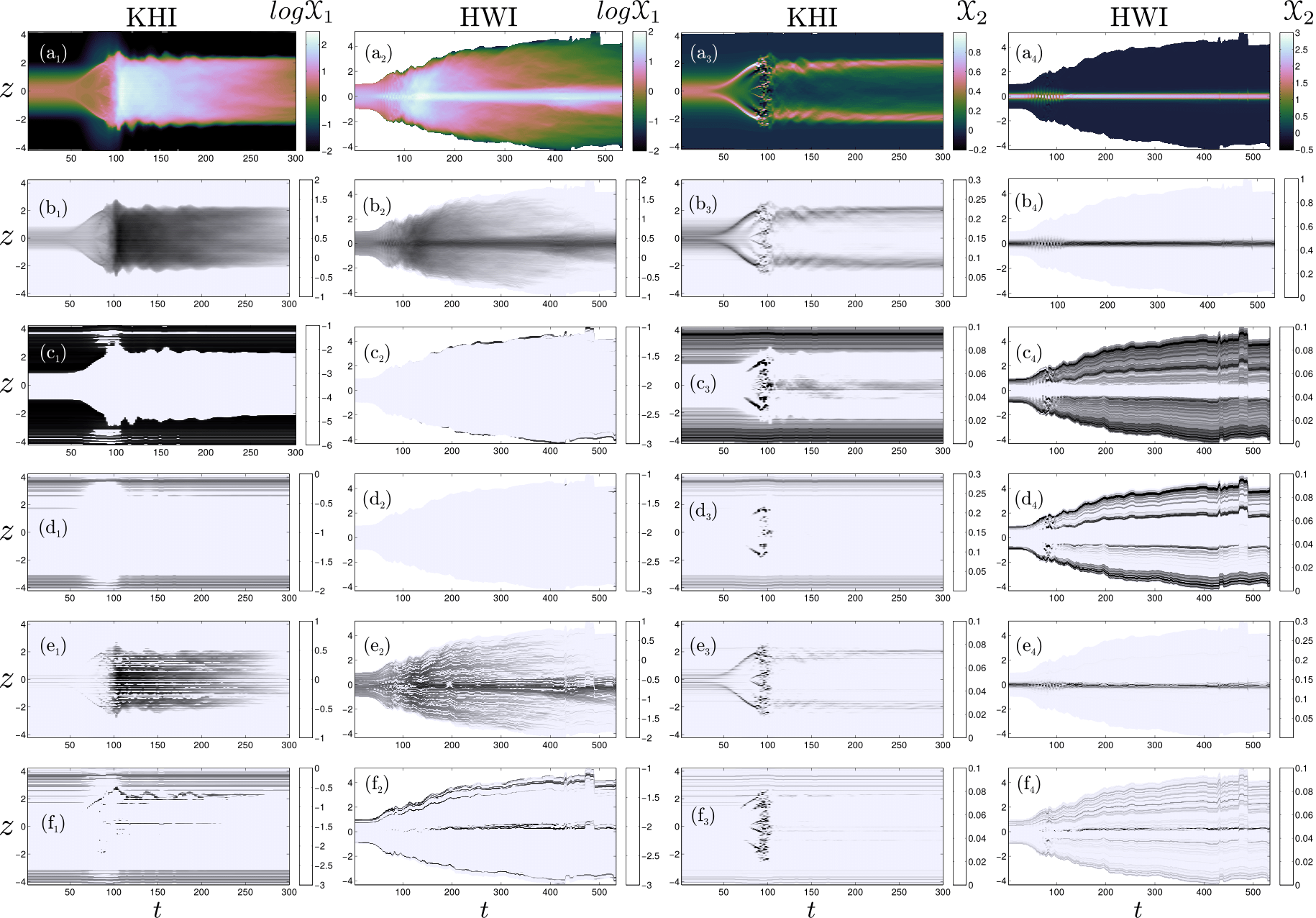}}
  \caption{($a_1$--$a_4$) Spatiotemporal structure of normalized  $N^2(z,t)$ and $\overline{\epsilon}(z,t)$ defined as $\mathcal{X}_1$ and $\mathcal{X}_2$ in \eqref{eq:inputs} (at a given instance $t_0$) for a representative KHI case (labeled as case (c) in figure \ref{fig:dl_eff}) and a representative HWI case (labeled as case (g) in figure \ref{fig:dl_eff}). The following rows include outputs of the first convolution layer in CNN6 that demonstrate filtered versions of their corresponding field (either $\mathcal{X}_1$ in log-scale or $\mathcal{X}_2$). In the text, these filters are referred to by the alphabetical part of their labels. For instance filter (b) produces panels ($b_1$--$b_2$) associated with $\mathcal{X}_1$ for the KHI and HWI cases respectively.
  }
\label{fig:conv_outputs}
\end{figure}

It is expected that the network learns the inherent (and intricate) patterns amongst the spatial structures of $N^2(z)$ and $\overline{\epsilon}(z)$ and maps it properly to the true DNS-based values of mixing efficiency. Figure \ref{fig:conv_outputs}($a_1$--$a_4$) illustrates the evolution of the local structures within these profiles for one KHI and one HWI case among the unseen validation set; more in-depth discussion on these structures are provided in \cite{SCP16} (see \eg their figure 12). The other panels in this figure illustrate the corresponding outputs of the first convolution layer that are filtered by various kernels whose weights and biases have been learned during the training procedure based on the CNN6 network (or CNN for brevity). For brevity and clarity, only the five most descriptive filtered outputs (out of 64), that have been hand-picked, are shown here which are denoted respectively as filters (b,c,d,e,f) as per their labels in figure \ref{fig:conv_outputs}. 

It appears that filter $b$ reproduces the structure of its input profiles merely at a different amplitude through \eg a simple linear scaling. Filters $c$ \& $d$, on the other hand, seem to produce an interesting attenuation of less important (insofar as mixing efficiency is concerned) segments of the profile. These segments include regions with negligible turbulent dissipation (see figure \ref{fig:conv_outputs} ($c_1$--$c_2$, $d_1$--$d_2$) or regions with $N^2(z,t) \approx 0$ (see figure \ref{fig:conv_outputs} ($c_3$--$c_4$, $d_3$--$d_4$). In contrast, filters $e$ \& $f$ have been trained to detect and isolate features of the input profiles that contribute more prominently to irreversible mixing. Figure \ref{fig:conv_outputs}($c_2$, $d_2$) illustrates zero output fields for HWI implying that our basic approach to isolate quiescent regions of the profile (discussed in \S\ref{sec:pre_process}) needs hardly any improvement for the HWI case, unlike that for the KHI case.
It is crucial to note that the identification of relevant regions with distinct dynamical effects on mixing has \emph{emerged} inevitably through the training procedure of our deep neural network and is surprisingly reminiscent of (at least qualitatively) the identification into quiescent (by filter c and d), intermittent (\eg by filter $f$) and turbulent patches (\eg by filter $e$) proposed by \cite{portwood2016} in the context of homogeneous stratified turbulence. We therefore believe a similar approach based on a convolutional neural network could be ideally suited to classify a turbulent field into these distinct regions.

As demonstrated in figure \ref{fig:conv_outputs}($a_1$--$a_4$), HWI and KHI have categorically different localization of $N^2(z)$ and $\overline{\epsilon}(z)$. It is nonetheless very interesting that a single convolution kernel, that has been trained with a disproportionately higher number of KHI examples, results in extracted features that are meaningful (and not distorted) for HWI, regardless of this difference in localization of these vertical profiles. In other words, the extracted features represent repeated patterns that are not tied to a specific position in the input field $\boldsymbol{\mathcal{X}}$. This might be explained by recalling that the CNN architecture has the important property of \emph{parameter sharing} that is inherent in the convolution operator. This property implies a strong prior knowledge that essentially assumes the nearby and local values of co-located $\epsilon(z)$ and $N^2(z)$ may have self-similar patterns that are relevant in approximating the induced mixing efficiency. Moreover, the pooling operator encourages the network to learn features that are translationally invariant. As a result, the CNN network is able to detect similar patterns even when the characteristic structure of the normalized  $\epsilon(z)$ and $N^2(z)$ are localized very differently; an issue that becomes particularly relevant to the two `atoms' of stratified turbulence investigated herein.

\begin{figure}
  \centerline{\includegraphics[width=.8\textwidth]{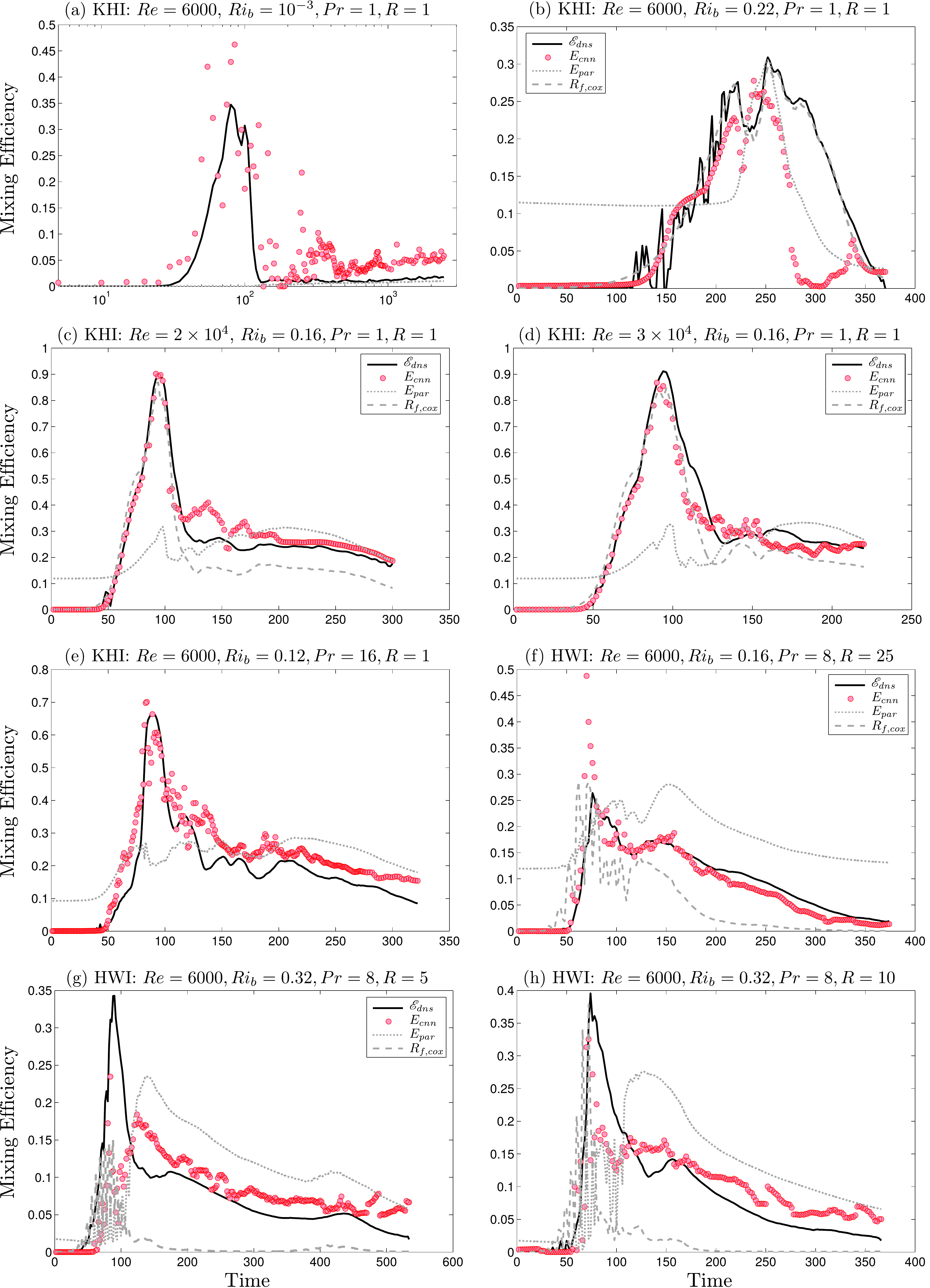}}
  \caption{Comparing the precise calculation of mixing efficiency, $\mathscr{E}_{dns}$ \eqref{eq:eff_dns} with those predicted by a convolutional neural network, denoted by $E_{cnn}$ (this study), and the most recent multi-parameter parameterization \citep{Hesam_GRL}, denoted by $E_{par}$, and that estimated by the Osborn-Cox method, $R_{f,cox}$ (defined in appendix \S\ref{sec:app:cox}, not available in panels (a,e)). Each panel illustrates temporal evolution of mixing efficiency due to either KHI or HWI under the various specified initial conditions for the validation set listed in Table \ref{tab:siminfo}.}
\label{fig:dl_eff}
\end{figure}


Next we assess to what extent the learned mapping function can be generalized to the \emph{unseen} validation dataset. Figure \ref{fig:dl_eff} demonstrates the prediction results of our deep learning approach based on the CNN network (see figure \ref{fig:cnn}), denoted by $E_{cnn}$ as well as their true DNS-based values (denoted by $\mathscr{E}_{dns}$). For further comparison, this figure also includes estimates of mixing efficiency due to (i) the Osborn-Cox method, denoted by $R_{f,cox}$ (refer to appendix \ref{sec:app:cox} for its definition), and (ii) the multi-parameter parameterization of \cite{Hesam_GRL} denoted by $E_{par}$. The latter relies on two dimensionless parameters, the buoyancy Reynolds number $Re_b(t)=\langle\overline{\epsilon}(z,t)\rangle/(\nu \langle N^2(z,t) \rangle)$ and a bulk Richardson number $Ri(t) = \langle N^2(z,t) \rangle/ \langle (d\overline{\boldsymbol{u}}(z,t)/dz)^2 \rangle$. Figure \ref{fig:dl_validationError}b provides the mean-squared error associated with these various estimates applied to each validation set, labeled as in figure \ref{fig:dl_eff}. 
As mentioned earlier in \S\ref{sec:dataset}, we have intentionally chosen the validation dataset to include simulations with extreme initial parameters (to avoid trivial `interpolation' between the training dataset). The associated results in figure \ref{fig:dl_eff} therefore consist of (a) very weakly and (b) very strongly stratified KHI, (c,d) KHI with extremely high initial values of $Re$, (e) KHI at high $Pr$, (f) HWI with a density layer that is very much sharper than its shear layer with $R=25$ and (g,h) very strongly stratified HWI.

Our CNN-based predictions are markedly superior to those predicted by the Osborn-Cox model or indeed by any published parameterization of mixing efficiency including our own most recent suggestion \citep{Hesam_GRL}. The predictions of $E_{cnn}$ are exceptionally accurate at higher Reynolds (c,d) and Prandtl (e) numbers considering that the ensuing turbulence is significantly more energetic than those employed for training purposes. Perhaps most surprising is the reasonable accuracy of $E_{cnn}$ for HWI-induced turbulence under strong stratification with $Ri_b=0.32$ (g,h). As discussed in depth in \cite{SOC_Hesam}, unlike KHI that is quite sensitive to its initial conditions, HWI reveals the striking characteristics that (regardless of its initial conditions) it self-organizes towards a critical state with a particular distribution of mean density and velocity (i.e. a critical state associated with a high probability density function of $Ri_g(z) = N^2(z)/ (d\overline{u}/dz)^2$ near 1/4). Furthermore, the mechanics involved in this self-organization are entirely different for a given $Ri_b$ or even depending on the thickness ratio $R$. Remarkably however, the universal common features discovered by the network reveals plausible transferability to HWI at significantly higher $Ri_b$ (g,h) or $R$ (f), as if the network has learned the pathways available for self-organization!  While the predictions of $E_{cnn}$ in (a) for KHI under extremely weak stratifications might have the largest variance compared to $\mathscr{E}_{dns}$, it is nonetheless very interesting that the increasing trend of $\mathscr{E}_{dns}$ with time is correctly predicted by CNN. The mixing efficiencies under such weak stratifications (\eg the training case with $Ri_b=5\times10^-3$) are so small that they do not impact adversely the mean-squared-error loss function employed during training. This may explain the higher variance of $E_{cnn}$ observed in case (a) despite its low MSE as shown in figure \ref{fig:dl_validationError}b. For strongly stratified KHI (panel b), CNN predicts accurately the evolution of mixing efficiency towards its maximum but suggests a more rapid decay than that inferred from $\mathscr{E}_{dns}$. The underlying reason for this relative inaccuracy of $E_{cnn}$ during a short period is not known to us.

Although $R_{f,cox}$ relies on additional information regarding the scalar dissipation field that is not required as an input by our deep neural network, its predictions are not consistently accurate. Most worrisome is perhaps for strongly stratified HWI (see panels (g,h)), where $R_{f,cox}$ predicts essentially negligible mixing, a prediction that is simply erroneous. For KHI cases $R_{f,cox}$ estimates are reasonable, albeit being less accurate than $E_{cnn}$ (see figure \ref{fig:dl_validationError}b) with the exception of case (b) where Osborn-Cox estimates are almost perfect. The parameterization of \cite{Hesam_GRL} has been constructed entirely based on the fully turbulent flows that are \emph{only} subject to KHI. As a result and as expected, $E_{par}$ systematically over-estimates the efficiency for HWI cases and fails to capture high efficiencies attained during the convectively unstable roll-up of primarily instabilities of either KHI or HWI type.


%
%
%
%



\section{Summary} 
\label{sec:summary} 

%
%
%
%
%
%

Using properly normalized vertical structures of $N^2$ and $\overline{\epsilon}(z)$, we have proposed a data-driven approach based on deep convolutional neural networks (CNN) that can accurately predict the value of mixing efficiency for the entire life cycle of KHI and HWI (i.e. two `atoms' of turbulence in stratified flows) beyond the range of initial conditions that have been employed for training the network. The large overturns of mixing that are convectively unstable are no longer ignored in such an approach. We have also shown that the results of the CNN model for KHI and HWI are more reliable and accurate than those based on the Osborn-Cox method.


Deep neural networks have a compositional hierarchy in which low-level features are composed to form higher-level features (\eg for image recognition the first layers of a CNN detect basic abstract features such as edges, then deeper layers combine edges to form motifs and subsequent layers assemble parts from motifs). We believe the proposed CNN model has similarly discovered such an `abstract' level of stratified turbulence with characteristics that are so universal that even with a small portion of data associated with HWI, the generic behavior of its induced mixing efficiency can be predicted robustly for wildly different initial conditions.

What makes such a data-driven approach especially appealing is its capability to become increasingly more accurate, robust and generic. This is foreseen to be achieved by (i) experimenting with many other types of DNN architectures, (ii) tuning the hyper-parameters (of which there are many) and perhaps most importantly (iii) further enriching the training dataset by adding additional examples of KHI and HWI, as they become available, or perhaps more excitingly by including more `atoms' of ocean turbulence such as those induced by \eg double-diffusion, Taylor and Rayleigh-Taylor instabilities. Another exciting future direction would involve using observed profiles (either from laboratory or real environments) to estimate mixing efficiency based on the proposed model, especially due to the relative inaccuracies of the Osborn-Cox method. 



\vspace{-0.35cm}
\section*{Acknowledgement}
\label{sec:acknowledgement}
H.S. acknowledges the SOSCIP TalentEdge postdoctoral fellowship and the support from Autodesk Research. H.S. is especially grateful to Dr. Francesco Iorio. 
The source codes, based on Google's TensorFlow library, are publicly available at \href{https://github.com/hsalehipour/dl-Osborn}{this github repository}. The research of WRP is provided by NSERC Discovery Grant A9627.

\vspace{-0.35cm}
\appendix
\section{}\label{sec:app:cox}
%
An alternative measure of mixing efficiency is the flux Richardson number, $R_f$, which assumes that the buoyancy flux, $\mathbb{B}$, is an appropriate quantity to describe diapycnal mixing $\mathscr{M}$:
\begin{equation}
 R_f(t) = \frac{\mathbb{B}(t)}{\mathbb{B}(t) + \langle \overline{\epsilon}(z,t) \rangle}.
 \label{eq:Rf}
\end{equation}
A widely used method for estimating mixing efficiency from observational profiles (see \eg \cite{Monismith_GRL_2018}) is that following \cite{Osborn_Cox_1972}. In this method $\mathbb{B}$ is estimated using the scalar dissipation rate $\chi = 2\kappa\langle \overline{|\nabla \rho'|^2} \rangle $ as:
\begin{equation}
 \mathbb{B}_{cox}(t) = \frac{\chi}{2}\langle N^2 \rangle \left({d\overline{\rho} \over dz} \right)^{-2},
\end{equation}
where turbulent fluctuations of the density field are defined as $\rho'(\boldsymbol{x},t) = \rho(\boldsymbol{x},t) - \overline{\rho}(z,t)$. Therefore $R_{f,cox}$ (using the `Cox' method), plotted in figure \ref{fig:dl_eff} based on the original DNS data, is computed by inserting $\mathbb{B}_{cox}$ into \eqref{eq:Rf}.

\bibliographystyle{jfm}
\bibliography{DL_eff}

\end{document}